\documentclass[useAMS]{mn2e}
\usepackage{epsf}

\newcommand{\xt}[1]{\mbox{$\times 10^{#1}$}}
\newcommand{\x}[1]{\hspace*{#1mm}}
\newcommand{\apj}{ApJ}
\newcommand{\apjs}{ApJS}
\newcommand{\aap}{A\&A}
\newcommand{\ctr}[1]{\multicolumn{1}{c}{#1}}
\def\nt#1{\vtop{\footnotesize\hsize=\columnwidth\leavevmode#1\hspace*{\fill}}}

\begin{document}

\title[Grain size distributions and photo-electric heating in ionized media]
{Grain size distributions and photo-electric heating in ionized media}

\author[P.A.M. van Hoof et al.] {P.A.M. van Hoof,$^{1,2,}$\thanks{E-mail:
p.van-hoof@qub.ac.uk}$^,$\thanks{Current address: Queen's University Belfast,
Physics Department, APS Division, Belfast, BT7~1NN, Northern Ireland.}
J.C.~Weingartner,$^{1,}$\thanks{Current address: The Department of Physics
and Astronomy, George Mason University, 4400 University Drive, MSN~3F3,
Fairfax, VA 22030, USA.}
P.G.~Martin,$^{1}$ K.~Volk,$^{3,}$\thanks{Current address: Gemini Observatory,
Southern Operations Center, Colina el Pina S/N, La Serena, Chile.} G.J.~Ferland$^{2}$\\
$^{1}$Canadian Institute for Theoretical Astrophysics, University
of Toronto, 60 St.\  George Street, Toronto, ON M5S~3H8, Canada\\
$^{2}$University of Kentucky, Dept.\  of Physics and Astronomy,
177 CP Building, Lexington, KY 40506--0055, USA\\
$^{3}$University of Calgary, 2500 University Dr. NW, Calgary, AB T2N~1N4, Canada}
\date{received, accepted}
\maketitle

\begin{abstract}
Ever since the pioneering study of Spitzer, it has been widely recognized that
grains play an important role in the heating and cooling of photo-ionized
environments. This includes the diffuse ISM, as well as H\,{\sc ii} regions,
planetary nebulae, and photo-dissociation regions. A detailed code is
necessary to model grains in a photo-ionized medium since the interactions of
grains with their environment include a host of microphysical processes. In
this paper we will use the spectral synthesis code Cloudy for this purpose. A
comprehensive upgrade of the grain model has been recently incorporated into
Cloudy. One of these upgrades is the newly developed hybrid grain charge
model. This model allows discrete charge states of very small grains to be
modelled accurately while simultaneously avoiding the overhead of fully
resolving the charge distribution of large grains, thus making the model both
accurate and computationally efficient. A comprehensive comparison with the
fully resolved charge state models of Weingartner \& Draine (2001a) shows that
the agreement is very satisfactory for realistic size distributions. The
effect of the grain size distribution on the line emission from photo-ionized
regions is studied by taking standard models for an H\,{\sc ii} region and a
planetary nebula and adding a dust component to the models with varying grain
size distributions. A comparison of the models shows that varying the size
distribution has a dramatic effect on the emitted spectrum. The strongest
enhancement is always found in optical/UV lines of the highest ionization
stages present in the spectrum (with factors up to 2.5 -- 4), while the
strongest decrease is typically found in optical/UV lines of low ionization
lines or infrared fine-structure lines of low/intermediate ionization stages
(with reductions up to 10 -- 25\%). Changing the grain size distribution also
affects the ionization balance, and can affect resonance lines which are
very sensitive to changes in the background opacity. All these results clearly
demonstrate that the grain size distribution is an important parameter in
photo-ionization models. 
\end{abstract}


\section{Introduction}

Grains are ubiquitous in the interstellar medium (ISM), and they can be
detected either directly through their far-infrared emission or indirectly
through extinction or polarisation studies. Despite the vast number of
observations, many questions regarding grain composition and grain physics
remain unanswered. Further study is therefore required, and detailed models
are needed to interpret the results. Ever since the pioneering study of
Spitzer (1948), it has been widely recognized that grains play an important
role in the heating and cooling of the diffuse ISM (see also the more recent
studies by Bakes \& Tielens 1994, and Weingartner \& Draine 2001a, hereafter
WD). Grains also play an important role in the physics of H\,{\sc ii} regions
and planetary nebulae (PNe; e.g., Maciel \& Pottasch 1982, Baldwin et al.\
1991, hereafter BFM, Borkowski \& Harrington 1991, Ercolano et al. 2003)
and photo-dissociation regions (PDR's; e.g., Tielens \& Hollenbach 1985).

The interactions of grains with their environment include a host of
microphysical processes, and their importance and effects can only be judged
by including all of these processes in a self-consistent manner. This
can, in turn, only be done with a complete simulation of the environment. In
this paper we use the spectral synthesis code Cloudy for this purpose. Cloudy
is a well known and widely used photo-ionization code. This code is not only
useful for modelling fully ionized regions, but calculations can also be
continued into the PDR. In order to make the models realistic, the presence of
a detailed grain model is usually required. The first grain model was
introduced into Cloudy in 1990 to facilitate more accurate modelling of the
Orion nebula (for a detailed description see BFM). In subsequent years, this
model has undergone some revisions and extensions, but remained largely the
same.

In the last couple of years, Cloudy has undergone several major upgrades,
described in Ferland (2000a), Ferland (2000b), and van Hoof et al.\ (2000b).
This includes a comprehensive upgrade of the grain model. The latter was
necessary for two reasons. First, the discovery of crystalline silicates in
stellar outflows (e.g., Waters et al.\ 1996), and other detailed observations
of grain emission features by the Infrared Space Observatory (ISO), meant that
the code had to become much more flexible to allow such materials to be
included in the modelling. Second, even before the ISO mission it had already
become clear that the photo-electric heating and collisional cooling of the
gas surrounding the grains is dominated by very small grains (possibly
consisting of polycyclic aromatic hydrocarbons or PAH's). The physics of very
small grains could not be modelled very accurately with the original grain
model. In view of these facts we have undertaken a comprehensive upgrade of
the grain model in Cloudy. The two main aims were to make the code more
flexible and versatile, and to make the modelling results more realistic
(e.g., by improving the treatment of grain charging, the photo-electric
effect, and stochastic heating). The new grain model has been introduced in
version 96 of Cloudy.

We have used a three-pronged approach to improve the grain model in Cloudy.
First we introduced a Mie code for spherical particles, which allows the user
to use arbitrary grain materials and resolve any grain size distribution of
their choosing to arbitrary precision. The latter is very important since most
grain properties depend strongly (and more importantly non-linearly) on size.
This upgrade, briefly described in Section~\ref{resolv}, also opened the
way for two other major improvements. First, it enabled the accurate modelling
of stochastic heating effects for arbitrary grain materials and size
distributions. Second, it enabled a much more realistic modelling of grain
charging, photo-electric heating, and collisional cooling by the grains, as
described in Section~\ref{physics}. For this purpose we have developed a
completely new grain charge model, which we call the hybrid model. This
is described in detail in Section~\ref{hybrid}.

In this paper we will study photo-electric heating by grains in photo-ionized
environments in detail. In particular, we will study the effect that the
distribution of grain sizes has on the relative intensities of emission lines.
We will show that this effect is nothing short of dramatic, making the grain
size distribution an important parameter in the modelling of
photo-ionized regions such as H\,{\sc ii} regions and planetary nebulae. This
will be described in Section~\ref{heating}. Our conclusions will be summarized
in Section~\ref{conclusions}.

\section{Resolving the grain size distribution}
\label{resolv}

\begin{figure}
\centerline{
\epsfxsize=0.9\columnwidth\epsfbox[64 316 493 653]{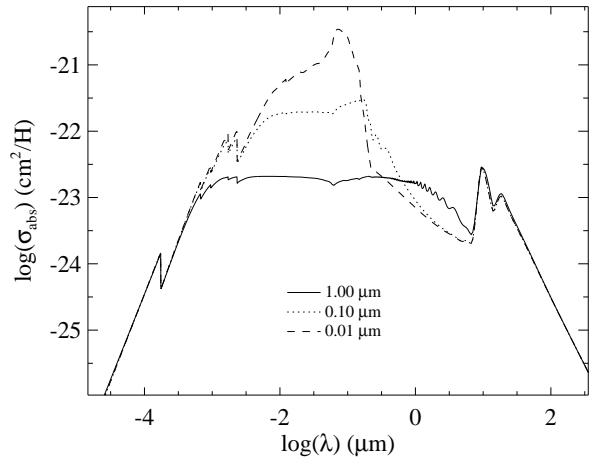}}
\caption{The absorption cross section for astronomical silicate (Martin \&
Rouleau 1991) for three single sized grains. The dust-to-gas ratio is the same
for all three species and the cross sections are normalized per hydrogen
nucleus in the plasma.
\label{plot:sd:sil}}
\end{figure}

\begin{figure}
\centerline{
\epsfxsize=0.9\columnwidth\epsfbox[64 316 493 653]{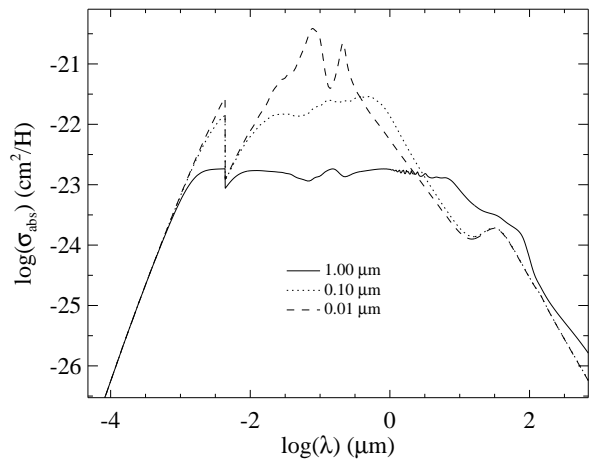}}
\caption{Same as Fig.~\ref{plot:sd:sil}, but for graphite (Martin \& Rouleau
1991).\label{plot:sd:gra}}
\end{figure}

In the original grain model of Cloudy, opacities for a handful of grain
species were hard-wired in the code. Furthermore, all grain properties would
be integrated or averaged over the entire size distribution. This is not a
very good approximation since most of these properties depend strongly on
size. This approach was nevertheless adhered to in BFM because of
computational restrictions.

Resolving the size distribution into many small bins improves the modelling in
several ways. First, the absorption cross sections of small grains are very
different from large grains (see Figs.~\ref{plot:sd:sil} and
\ref{plot:sd:gra}). Resolving the size distribution into many size bins
enables the equilibrium temperature to be calculated for each bin separately.
This leads to a more accurate prediction for the spectrum since grain
emissions are a strongly non-linear function of temperature.

More importantly, resolving the size distribution also enables other
improvements of the grain treatment: stochastic heating can be treated
correctly for the smallest grains in the size distribution (this will be
discussed in more detail in a forthcoming paper), and the calculations yield
much more accurate results for grain charging, photo-electric heating and
collisional cooling of the gas by the grains (see
Section~\ref{physics}).

To improve the model, we have implemented the following changes:

1 -- We have included a Mie code for spherical particles in Cloudy. Assuming
that the grains are homogeneous spheres with a given complex refractive index
(optical constant) one can use Mie theory (Mie 1908) to calculate the
absorption and scattering opacity. This has to be done separately for every
wavelength since the refractive index depends on wavelength. Good overviews of
Mie theory can be found in van de Hulst (1957), and Bohren \& Huffman (1983).
Our Mie code is based on the program outlined in Hansen \& Travis (1974) and
references therein. The optical constants needed to run the code are read from
a separate file. This allows greater freedom in the choice of grain species.
Files with optical constants for a range of materials are included in the
Cloudy distribution. However, the user can also supply optical constants for a
completely different grain type.

2 -- Several mixing laws have been included in the code (Bruggeman
1935; Stognienko et al. 1995; and Voshchinnikov \& Mathis 1999, based on
theory described in Farafonov 2000). This allows the user to define grains
which are mixtures of different materials. Cloudy will then calculate the
appropriate opacities by combining the optical constants of these grain types.

3 -- It is possible to use arbitrary grain size distributions. The user can
either choose one of a range of preset functions (with numerous free
parameters), or supply the size distribution in the form of a table.
Single-sized grains can also be treated.

4 -- The size distribution can be resolved in an arbitrary number of size bins
(set by the user), and the absorption and scattering opacities and all the
physical parameters (charge, temperature, etc.) are calculated for each bin
separately.

\section{Changes to the grain physics}
\label{physics}

We have modified certain aspects of the grain physics following the discussion
in WD. Below we highlight certain aspects of these changes. A detailed
description will be presented in a forthcoming paper.

1 -- We include the bandgap between the valence and conduction bands in our
potential well model for silicates. This change only affects the results for
negatively charged grains ($Z_{\rm g} \leq -1$).

2 -- Reduction of the potential barrier for negatively charged grains is
included using an analytic fit to numerical calculations. Two effects are
important here: quantum tunnelling and the Schottky effect. Quantum theory
predicts that an electron with insufficient energy to overcome a barrier still
has a finite chance of tunnelling through. This effect has been modelled using
the WKB approximation which gives a simple analytic expression for the
tunnelling probability for a barrier of given width and height. Quantum
tunnelling is only important for small grains. For large grains the Schottky
effect will dominate, which describes the lowering of the potential barrier by
an image potential in the grain. This effect has been accurately modelled by
Draine \& Sutin (1987).

We will approximate both effects by assuming that the barrier is effectively
reduced in height from \mbox{$-(Z_{\rm g}+1)e^2/(4\pi\epsilon_0 a)$} to
$-E_{\rm min}$. The magnitude of the combined tunnelling/Schottky effect was
calculated by WD. However, the fitting function they used has the wrong
limiting behaviour for large grains where it should asymptotically approach the
classical Schottky expression. We therefore repeated these calculations using
the same assumptions, but adopted a different fitting function that does
exhibit the correct limiting behaviour:
\begin{equation}
\label{thres:new}
    E_{\rm min} = -  \, \theta_\nu \frac{e^2}{4\pi\epsilon_0 a}
    \left[ 1  - \frac{0.3}{ ( a / {\rm nm} )^{0.45} \, \nu^{0.26}} \right],
\end{equation}
where $\theta_\nu[\nu = -(Z_{\rm g}+1) \, ]$ describes the Schottky
effect and is defined in Draine \& Sutin (1987). The term in square brackets
describes the quantum-mechanical correction. This change only affects the
results for grains with $Z_{\rm g} < -1$.

3 -- The treatment of the photo-electric effect has been improved, following
the discussion in WD. This includes new expressions for the ionization
potential, photo-electric yield, and the energy distribution of ejected
electrons.

4 -- Certain physical constants have been updated. Most notably, the work
function for graphite has been lowered. The old value was equal to that
of silicate, which was unrealistically high. The change results in an
increased photo-electric heating rate.

5 -- The treatment of electron sticking probabilities has been updated,
again following WD. Especially for very small grains the sticking efficiency
has been substantially lowered to obtain better agreement with laboratory
studies of molecules. This has an important impact on the photo-electric
heating rate of the gas since the electron recombination rate has to be
matched by electron loss processes to preserve the charge balance. The loss
processes are usually dominated by the photo-electric effect. We also
introduced a minimum charge for the grains, as outlined in WD. This
modification is relevant for very small grains in fully molecular regions.
This change can have an important effect on the amount of free electrons, as
well as on the amount of heating from photo-detachment.

6 -- The treatment of collisional processes between charged particles and
the grains have been improved. The modification factors for Coulomb
attraction or repulsion of incoming particles have been upgraded following
Draine \& Sutin (1987) who include image potential effects in the grains. We
have also modified the physics for charge exchange between ions and grains.
The new code is based on the assumption that electrons move into the deepest
potential well, either the grain or the ion. In certain circumstances this may
be different from the old assumption that ions always recombine to their
neutral state upon impact. This change has little direct impact on the heating
and cooling rates, but it can influence the grain charging and the ionization
balance in the gas. In turn this can influence photo-electric heating rates.

Our treatment deviates from the WD code in two ways. Most importantly, we use
a different grain charge model, which will be discussed in more detail in
Section~\ref{hybrid}. Secondly, we use slightly different physics for 
charge exchange between ions and grains, as outlined above in point 6.
The latter only gives rise to very small differences at the 1 -- 2\%
level or better when compared to the WD treatment.

\subsection{The new hybrid grain charge model}
\label{hybrid}

The original grain model in Cloudy (which we will call the average grain
potential model) is described in BFM. In that model an average grain potential
is calculated by finding the potential for which the charge gain rate exactly
matches the loss rate. This method was first proposed by Spitzer (1948), and
is an excellent approximation for large grains. However, it is now clear that
photo-electric heating and collisional cooling of the gas are dominated by
very small grains. For such grains the average grain potential approximation
does not work very well because grain physics becomes increasingly non-linear
as a function of charge for smaller grain sizes. This fact, combined with the
fact that grain charges are quantized, has led to a new approach where the
charge distribution is fully resolved, and heating and cooling rates are
calculated for each charge state separately (see e.g., WD). This ensures
accurate results, but leads to an appreciable increase in computational
overhead. This is especially the case for large grains since the width of the
charge distribution increases with grain size. Hence the paradoxical situation
arises that most of the computing time is spent on grains which contribute
least to heating and cooling, and which are also the grains for which the
average grain potential model works best!

In this paper we present a hybrid grain potential model which is almost
as computationally efficient as the original average grain potential model,
but nevertheless gives sufficient accuracy when compared to fully resolved
charge distribution calculations. The basic philosophy is that for very small
grains ($a < 1$~nm) only a few charge states have a significant population.
Hence we adopt the $n$-charge state approximation, in which all grains are
treated by using exactly $n$ contiguous charge states, independent of size.
The higher $n$ is, the more accurate the results will be (exactly how accurate
will be discussed in Section~\ref{validating}). The default for Cloudy
calculations is $n = 2$, but the user can request a larger number if higher
precision is desired.

Since the $n$-charge state model does not fully resolve the charge
distribution, a different algorithm from WD is needed to calculate the charge
states $Z_i \equiv Z_1 + i -1$, and the population of these states. The value
of the lowest of the $n$ grain charges, $Z_1$, is found using an iterative
procedure, as discussed below. The populations $f_i$ of the charge states must
first of all obey the following normalisation:
\begin{equation}
\label{c1}
   \sum_{i=1}^n f_i = 1.
\end{equation}
Secondly, we require that the electron gain rates $J_i^-$ and electron loss
rates $J_i^+$ summed over all charge levels match exactly:
\begin{equation}
\label{c2}
   \sum_{i=1}^n f_i ( J_i^+ - J_i^- ) = 0,
\end{equation}
similar to the average grain potential model. $Z_1$ is defined by
requiring that $J_i^+ - J_i^-$ changes sign between $Z_1$ and $Z_2$ when $n=2$.
Equations~\ref{c1} and \ref{c2} are sufficient to determine the charge state
populations if $n = 2$, but for $n > 2$ we need additional equations. These
equations need to satisfy the following constraints. First, the resulting
level populations should always be greater or equal to zero. Second, the level
populations should change continuously when the electron gain and loss rates
change continuously. Third, the level populations should asymptotically
approach the results from fully resolved calculations for increasing values of
$n$. We have adopted the following algorithm:

1 -- The $n$ charge states are split up in two groups of $n-1$ contiguous
charge states. The first group contains charge states $[Z_1,Z_{n-1}]$, and the
second $[Z_2,Z_n]$. The value for $Z_1$ is determined iteratively (see step~4).

2 -- The relative level populations in the first group $f_i^1$ are determined
using an algorithm very similar to the one used in fully resolved
calculations, i.e., assume $f_1^1 =1$, calculate $f_2^1 = f_1^1 J_1^+ / J_2^-$,
$f_3^1 = f_2^1 J_2^+ / J_3^-$, etc.\footnote{Note that this procedure is not
correct for charge transfer with multiply charged ions. In order to avoid
having to solve a full set of linear equations, we will approximate this
process as multiple single-charge-transfer events. The resulting errors are
expected to be small as collision rates for multiply charged ions are usually
quite low because their velocities are small compared to electrons. The
collision rates are normally suppressed even further by the positive grain
charge.}, and then re-normalise to $\sum_{i=1}^{n-1} f_i^1 = 1$. We use an
analogous procedure for the populations $f_i^2$ of the second group for $i \in
[2,n]$.

3 -- Determine for both groups the net charging rate
\begin{equation}
\label{c3a}
   J_k = \sum_{i=k}^{n-2+k} f_i^k ( J_i^+ - J_i^- ) \x{3} (k = 1,2).
\end{equation}

4 -- Iterate $Z_1$ and repeat steps 1 -- 3 until $J_1 \times J_2 \leq 0$. Then
find $0 \leq \alpha \leq 1$ such that
\begin{equation}
\label{c3b}
   \alpha J_1 + (1 - \alpha) J_2 = 0.
\end{equation}

5 -- Determine the final charge state populations as follows:
\begin{equation}
\label{c3c}
   f_i = \alpha \, f_i^1 + (1 - \alpha) \, f_i^2 \x{3} (\mbox{with } f_n^1 \equiv 0, f_1^2 \equiv 0).
\end{equation}
One can verify that this algorithm satisfies all constraints.

The hybrid grain potential model is efficient because it avoids the overhead
for large grains, while still giving accurate results for both small and large
grains. An added bonus is that most of the time an excellent initial estimate
for $Z_1$ can be derived from the previous zone, reducing the overhead even
further. The model works for very small grains because only few charge states
are populated and it can reconstruct the actual charge distribution. It works
for large grains because the grain potential distribution approaches a delta
function for increasing grain size (as opposed to the charge distribution
which becomes ever wider). Our method therefore asymptotically approaches the
average grain potential model, which we already know is very accurate for
large grains. No simple predictions can be made as to how the hybrid
grain charge model will behave for intermediate grain sizes. We therefore
conducted comprehensive tests which will be discussed in
Sect.~\ref{validating}.

\subsection{Validating the hybrid grain charge model}
\label{validating}

\begin{table}
\small
\begin{center}
\caption{\leftskip 13mm\setlength\hsize{121mm}
Physical parameters for the benchmark models. Symbols have their usual
meaning, $G$ is the intensity of the radiation field and $G_0 =
1.6\xt{-6}$~W\,m$^{-2}$, integrated between 6 and 13.6~eV, is the Habing
intensity. $T_{\rm c}$ is the colour temperature of the radiation field.
\label{summ}}
\begin{tabular}{lrrrrrr}
\hline
 & \multicolumn{2}{c}{ISM} & \multicolumn{2}{c}{H\,{\sc ii}} & \multicolumn{2}{c}{PN} \\
 & warm & cold & ionized & PDR & ionized & PDR \\
\hline
$T_{\rm c}$/kK             &  35 &   35 &   50 & 50 & 250 & 250 \\
log($G/G_0$)               &   0 &    0 &    5 &  5 &   5 &   5 \\
log($n_{\rm H}$/cm$^{-3}$)\hspace{-5mm} &   0 &    1 &    4 &  4 &   4 &   4 \\
log($n_{\rm e}$/cm$^{-3}$)\hspace{-5mm} &   0 & $-$2 &    4 &  1 &   4 &   1 \\
$T_{\rm e}$/kK             &   9 &  0.1 &    9 &  1 &  20 &   1 \\
\hline
\end{tabular}
\end{center}
\end{table}

\begin{table}
\begin{center}
\caption{Summary of the comparison of photo-electric heating rates (top
panel) and collisional cooling rates (bottom panel) between the Cloudy
$n$-charge state calculations (indicated by CLD$n$) and the benchmark
calculations with the WD code. All entries are differences CLD$n$/WD $-$ 1 in
percent.\label{summary}}
\begin{tabular}{lrrrr}
\hline
 & \multicolumn{4}{c}{heating} \\
 & \multicolumn{2}{c}{single size} & \multicolumn{2}{c}{size distr.} \\
 & median & worst & median & worst \\
\hline
CLD2 & $-3.03$ & $-55.5$ & $-9.85$ & $-23.1$ \\
CLD3 & $-2.44$ & $-29.9$ & $-8.34$ & $-16.3$ \\
CLD4 & $-1.40$ &  $-9.6$ & $-3.43$ &  $-4.7$ \\
CLD5 & $-0.75$ &  $-5.9$ & $-1.45$ &  $-2.8$ \\
\hline
\end{tabular}
\vspace{3mm}
\begin{tabular}{lrrrr}
\hline
 & \multicolumn{4}{c}{cooling} \\
 & \multicolumn{2}{c}{single size} & \multicolumn{2}{c}{size distr.} \\
 & median & worst & median & worst \\
\hline
CLD2 & $-0.77$ & $-23.1$ & $-3.05$ & $-3.7$ \\
CLD3 & $-0.72$ & $-12.5$ & $-2.27$ & $-3.5$ \\
CLD4 & $-0.65$ &  $+5.2$ & $-1.30$ & $-3.1$ \\
CLD5 & $-0.44$ &  $+3.5$ & $-1.21$ & $-2.7$ \\
\hline
\end{tabular}
\end{center}
\end{table}

In order to validate the new grain charge model, we calculated the
photo-electric heating and collisional cooling rates for a range of physical
conditions, two grain species, and a wide range of grain sizes (including a
realistic size distribution). We modelled conditions typical for the warm and
cold ISM, H\,{\sc ii} regions (both the ionized region and the PDR surrounding
it), and planetary nebulae (again both the ionized region and the PDR). We
modelled the physical conditions with simple assumptions: the plasma only
contained hydrogen, the electron temperature and density were fixed at
prescribed values, and the incident spectrum was assumed to be a blackbody
(either full in the warm ISM and ionized cases, or cut off at 13.6~eV in the
cold ISM and PDR cases). The physical parameters are summarized in
Table~\ref{summ}. The grain materials were assumed to be astronomical silicate
and graphite (Draine \& Lee 1984), and the grain size distribution was the A6
case with $R_V$ = 3.1 taken from Weingartner \& Draine (2001b). We then
compared these calculations with benchmark results from the WD code, which
fully resolves the charge distribution. A detailed discussion of these tests
(including tables of photo-electric heating and collisional cooling rates) can
be found in van Hoof et al. \cite{vh01}.

In Table~\ref{summary} we summarise the comparison of the photo-electric
heating and collisional cooling rates from van Hoof et al. (2001). These
tables show the relative discrepancy between the WD and Cloudy results in
percent. In general the results are in excellent agreement, with only a few
outliers for single sized grains in the $n=2$ and $n=3$ cases. The origin
of these discrepancies is studied in more detail for the worst performing
single sized grain in both the $n=2$ and $n=3$ case: a 5~\AA\ silicate grain
in cold ISM conditions. The charge state populations from the WD and Cloudy
models are compared in the top panel of Table~\ref{worst}. The $n$-charge
state model was designed to find the charge distribution, and hence one would
expect that the average charge should be in reasonable agreement with fully
resolved calculations. The middle panel of Table~\ref{worst} shows that this
is indeed the case. Key to understanding the discrepancy in the photo-electric
heating rates is the observation that these rates behave non-linearly as a
function of charge, and that the rates are highest for the lowest charge
states. The non-linearity is strongest for very small grains close to their
lowest allowed charge state ($Z = -1$ for the 5~\AA\ silicate grain). This is
illustrated in the bottom panel of Table~\ref{worst}: the $Z = -1$ state
produces more than 62\% of the photo-electric heating, while less than 13\% of
the grains are in that charge state. Since the $n=2$ calculations produce an
average charge slightly above zero, the model is missing the $Z = -1$ state
which would have given the largest contribution to the photo-electric heating.
This also explains why the photo-electric heating rates from the $n$-charge
state model are consistently lower than the WD results, although they do
asymptotically approach the correct result for larger grain size or larger
$n$. When $n$ is not high enough to fully resolve the charge state
distribution, the photo-electric heating from the lowest charge states will be
missed. Even the increased population of some of the higher states cannot
fully make up for that loss and the total amount of heating will be
somewhat underestimated.

The results for the size distribution cases always agree to better than 25\%,
even for $n=2$. The worst-case performance is found in PDR type
conditions where grains tend to be negative, while in ionized regions the
results agree to better than 8\%. This is well within the accuracy with which
we know grain physics to date. There are still major uncertainties in the
photo-electric yields and the sticking efficiency for electrons, both of which
have a strong effect on the photo-electric heating rate. Also the work
function and bandgap for astrophysical grain materials are poorly known and
can have a strong effect as well. This is unfortunate since photo-electric
heating and collisional cooling in photo-ionized environments are important
effects. These uncertainties largely stem from our uncertain knowledge of the
composition of interstellar grains.

\begin{table}
\begin{center}
\caption{Comparison of the fractional charge state populations (top
panel), average charge (middle panel), and photo-electric heating rates
(bottom panel) of a 5~\AA\ silicate grain in cold ISM conditions resulting
from the WD and $n$-charge state calculations. The results from the 5-charge
state calculations are omitted since they are virtually identical to the
4-charge state results. Entries $a(-b)$ stand for $a$\xt{-b}.\label{worst}}
\begin{tabular}{rrrrr}
\hline
    & \multicolumn{4}{c}{fractional charge state populations} \\
$Z/{\rm e}$ & WD & CLD2 & CLD3 & CLD4 \\
\hline
 $-1$ & 0.1271 &    --- & 0.0594 & 0.1272     \\
  $0$ & 0.8325 & 0.9876 & 0.9152 & 0.8326     \\
  $1$ & 0.0404 & 0.0124 & 0.0254 & 0.0402     \\
  $2$ & 9.94($-6$) & --- &   --- & 4.97($-9$) \\
\hline
\end{tabular}
\vspace{3mm}
\begin{tabular}{rrrrr}
\hline
    & \multicolumn{4}{c}{average charge in e} \\
    & WD & CLD2 & CLD3 & CLD4 \\
\hline
$<\!\!Z\!\!>$ & $-0.0866$ & $+0.0124$ & $-0.0340$ & $-0.0870$ \\
\hline
\end{tabular}
\vspace{3mm}
\begin{tabular}{rrrrr}
\hline
    & \multicolumn{4}{c}{Photo-electric heating rates in W\,m$^{-3}$} \\
$Z/{\rm e}$ & WD & CLD2 & CLD3 & CLD4 \\
\hline
 $-1$ & 2.51($-27$) &         --- & 1.17($-27$) & 2.50($-27$) \\
  $0$ & 1.53($-27$) & 1.80($-27$) & 1.66($-27$) & 1.51($-27$) \\
  $1$ & 1.27($-31$) & 3.61($-32$) & 7.40($-32$) & 1.17($-31$) \\
  $2$ & 0.00\phantom{($-31$)} &         --- &         --- & 0.00\phantom{($-31$)} \\
\hline
\end{tabular}
\end{center}
\end{table}

From the bottom panel of Table~\ref{summary} one can see that the
collisional cooling rates usually are in better agreement than the
photo-electric heating rates for a given set of physical parameters. It is
furthermore clear that the accuracy of the $n$-charge state approximation
increases as $n$ increases, as should be expected. Closer inspection of
Tables~2, 3, and 4 from van Hoof et al.~(2001) for single sized grains reveals
that the largest errors in the $n=2$ and $n=3$ cases are for the 0.5~nm
grains, for $n=4$ for either 2~nm or 10~nm grains, and for $n=5$ for 10~nm
grains, i.e., the grain size for which the errors are largest shifts upwards
for higher values of $n$. This is expected since the $n$-charge state
approximation will fully resolve the charge distribution of the smallest
grains for $n > 3$. Note that the results for the 100~nm grains are always in
excellent agreement, even when $n=2$, despite the fact that the actual charge
distribution is much wider than that.

The agreement between the Cloudy and the WD results is very satisfactory for
realistic size distributions, and should be sufficient for all realistic
astrophysical applications. Therefore the hybrid grain charge model presented
above (with $n=2$) will be the default for Cloudy modelling. By issuing a
simple command, the user can choose a higher value for $n$ if higher precision
is desired. The default value for $n$ will be increased in the near future
when greater computer speed and/or greater efficiency of the algorithm will
allow us to do so. We will use $n=2$ below for a more detailed study of the
photo-electric heating effect in dusty ionized plasmas.

\section{Photo-electric heating in ionized regions}
\label{heating}

\subsection{Introduction}

\begin{table}
\begin{center}
\caption{Properties of the grains included in the models. For size
distributions where no distinction is made between the silicate and the
graphite component (single sized grains, MRN, and KMH), the entry is equally
valid for both materials. The total dust-to-gas mass ratio is 6.34\xt{-3} for
the H\,{\sc ii}-region models, 3.96\xt{-3} for the PN models containing
silicate grains, and 2.34\xt{-3} for the PN models containing
graphite.\label{grain:prop}}
\begin{tabular}{llrrrrrrrrrrr}
\hline
no. & label & $R_V$ & $a_{\rm min}$ & $a_{\rm max}$ & $n_{\rm bin}$ \\
    &       &       &   $\mu$m      &   $\mu$m      &               \\
\hline
\multicolumn{6}{c}{H\,{\sc ii} region models} \\
\hline
0 & no dust \\
1 & 1.0~$\mu$m        & --- & 1.00000 & 1.0000 &  1 \\
2 & 0.1~$\mu$m        & --- & 0.10000 & 0.1000 &  1 \\
3 & MRN55$^a$         & 5.5 & 0.03000\rlap{$^d$} & 0.2500 &  9 \\
4 & KMH53$^b$         & 5.3 & 0.00250 & 3.0000 & 31 \\
5 & WD55 A0 (sil)$^c$ & 5.5 & 0.00035 & 0.4186 & 31 \\
  & WD55 A0 (gra)$^c$ & 5.5 & 0.00035 & 4.2102 & 41 \\
6 & WD55 A3 (sil)$^c$ & 5.5 & 0.00035 & 0.4145 & 31 \\
  & WD55 A3 (gra)$^c$ & 5.5 & 0.00035 & 1.6662 & 62 \\
\hline
\multicolumn{6}{c}{Planetary nebula models (silicate)} \\
\hline
0 & no dust \\
1 & 1.0~$\mu$m        & --- & 1.00000 & 1.0000 &  1 \\
2 & 0.1~$\mu$m        & --- & 0.10000 & 0.1000 &  1 \\
3 & WD31 A0$^c$       & 3.1 & 0.00035 & 0.3852 & 30 \\
4 & WD31 A6$^c$       & 3.1 & 0.00035 & 0.3805 & 30 \\
5 & MRN31$^a$         & 3.1 & 0.00500 & 0.2500 & 17 \\
6 & KMH31$^b$         & 3.1 & 0.00250 & 3.0000 & 31 \\
\hline
\multicolumn{6}{c}{Planetary nebula models (graphite)} \\
\hline
0 & no dust \\
1 & 1.0~$\mu$m        & --- & 1.00000 & 1.0000 &  1 \\
2 & 0.1~$\mu$m        & --- & 0.10000 & 0.1000 &  1 \\
3 & MRN31$^a$         & 3.1 & 0.00500 & 0.2500 & 17 \\
4 & WD31 A0$^c$       & 3.1 & 0.00035 & 1.3399 & 38 \\
5 & KMH31$^b$         & 3.1 & 0.00250 & 3.0000 & 31 \\
6 & WD31 A6$^c$       & 3.1 & 0.00035 & 1.0222 & 60 \\
\hline
\end{tabular}
\end{center}
\nt{$^a$Mathis et al. (1977). $^b$Kim et al. (1994). $^c$Weingartner \& Draine
(2001b). $^d$The Mathis et al. (1977) size distribution was truncated at a
lower limit of 0.03~$\mu$m to simulate the grain size distribution in Orion.
See also the discussion in BFM.}
\end{table}

An energetic electron can be ejected from a grain following the
absorption of a photon. The photoelectrons share their energy with the gas via
collisions, raising the local electron temperature; this process is known as
photoelectric heating. This excess energy can then be used to enhance
collisional excitation of certain (usually forbidden) emission lines, which
will change the emitted spectrum. This effect is known and has been described
in Dopita \& Sutherland (2000) and Volk (2001). The photo-electric effect will
also alter the ionization balance in the plasma. In a typical photo-ionized
plasma the free electrons will not have enough energy for collisional
ionization to be an important process. However, the raise in electron
temperature will reduce the recombination rates and thereby lead to an overall
increase in ionization. All these processes are known. However, it is not
widely known that the size distribution of the grains plays a very important
role in determining the magnitude of this effect.

\subsection{Description of the models}

In order to test this, we have constructed a set of models with Cloudy 96
beta~5 based on the standard Paris H\,{\sc ii} region and planetary nebula
models (HII40 and PN150, resp., P\'equignot et al. 2001). The Paris
H\,{\sc ii} region model is valid for low-excitation photo-ionized gas, while
the Paris PN model is valid for high-excitation gas. The base models contain
no dust, and will be used as a point of reference. We constructed 6 models out
of each base model by simply adding a dust component. Two models adopt single
sized grains of 1.0 and 0.1~$\mu$m, while the other 4 adopt more or less
realistic size distributions taken from the literature, as indicated in
Table~\ref{grain:prop}. These size distributions were all constructed for the
purpose of reproducing interstellar extinction curves. They were used in our
models because they are the most detailed studies of grain size distributions
that can be found in the literature. In order
to keep the models plausible, we used size distributions that reproduce the
$R_V$ = 5.5 extinction curve for the H\,{\sc ii} region models, and size
distributions that reproduce the $R_V$ = 3.1 extinction curve for the PN
models. For the H\,{\sc ii} region model we used a mixture of silicates and
graphite (Martin \& Rouleau 1991) with a dust-to-gas mass ratio of
6.34\xt{-3}. In the planetary nebula cases, we made separate models for either
silicate or graphite (with a dust-to-gas mass ratio of 3.96\xt{-3} and
2.34\xt{-3}, resp.) as these two species are not expected to coexist spatially
in the nebular material on theoretical grounds\footnote{Note that spectroscopy
of planetary nebulae by ISO has revealed a surprisingly large number of cases
that show both silicate and graphitic dust features. One famous example is
NGC~6302 (Molster et al.\ 2001). It is however usually assumed that the two
species reside in different parts of the nebula.}. Note that the size
distributions presented in Weingartner \& Draine (2001b) were derived by
matching the extinction curve using grain opacities defined in Li \& Draine
(2001). In this study we use the same size distribution, but use graphite
opacities from Martin \& Rouleau (1991) instead, which is not consistent. This
inconsistency is irrelevant for our purposes; more detailed studies will have
to await the construction of realistic size distributions for H\,{\sc ii}
regions and planetary nebulae. In all cases the size distribution was resolved
to a precision $\ln(a_{n+1}/a_n) \approx 0.23$. This resolution is sufficient
to accurately converge the effects of the grains. The resulting number of size
bins, together with the lower and upper limit of the size range, is indicated
in Table~\ref{grain:prop}. We stress that in all comparisons the chemical
composition and the dust-to-gas mass ratio of the dust is the same, and the
only difference is the size distribution. We also point out that the models
are ionization bounded, so the outer radius varies, depending on the total
opacity of the grains (which also depends strongly on the size distribution).
Since the grains and the atomic gas are competing for ionizing photons, this
also creates a mild dependence of the overall ionization structure on the size
distribution of the grains.

\subsection{Results and Discussion}

\begin{figure*}
\centerline{\hfill
\epsfxsize=0.3\textwidth\epsfbox[120 318 450 653]{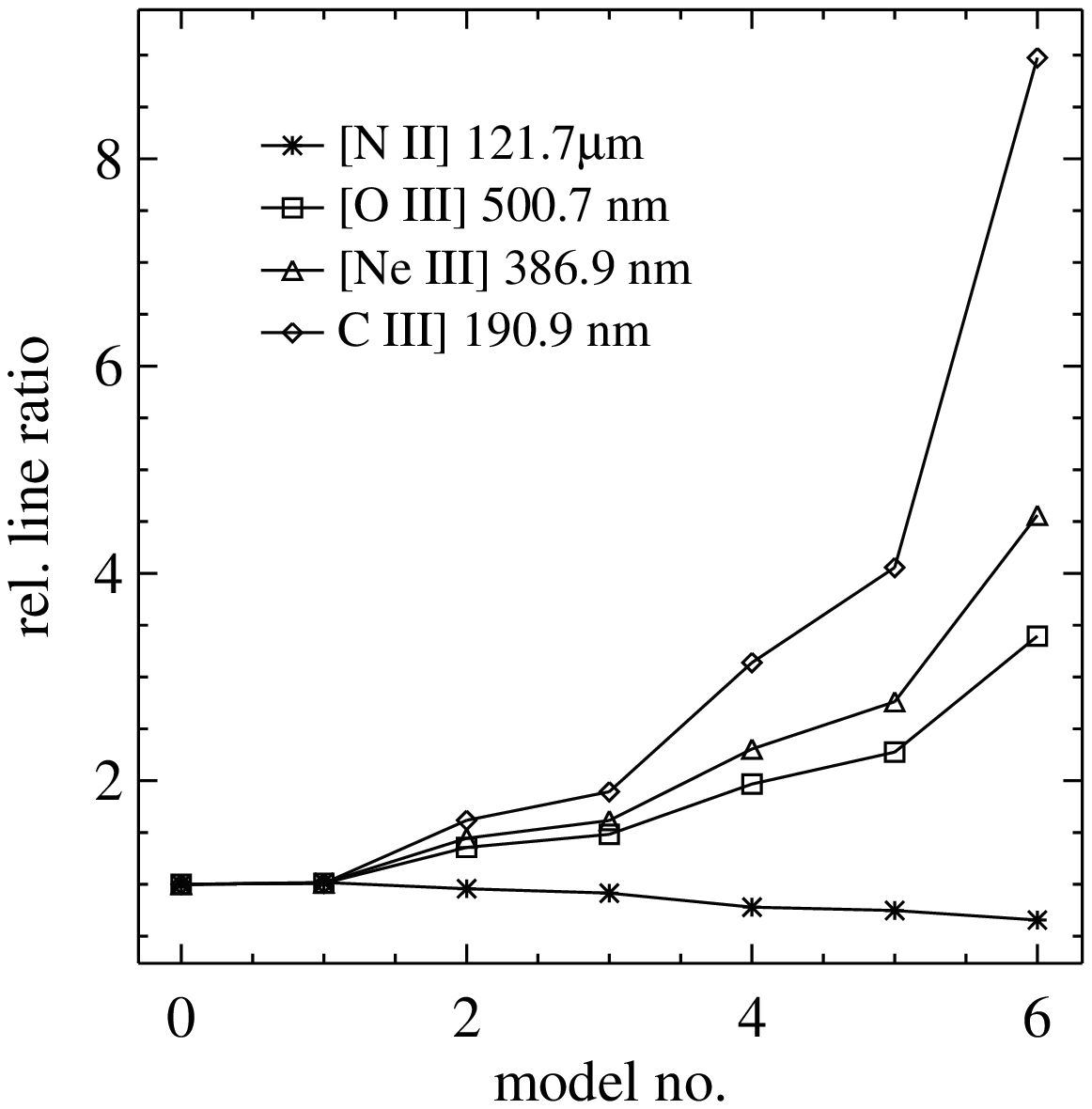}\hfill
\epsfxsize=0.3\textwidth\epsfbox[120 318 450 653]{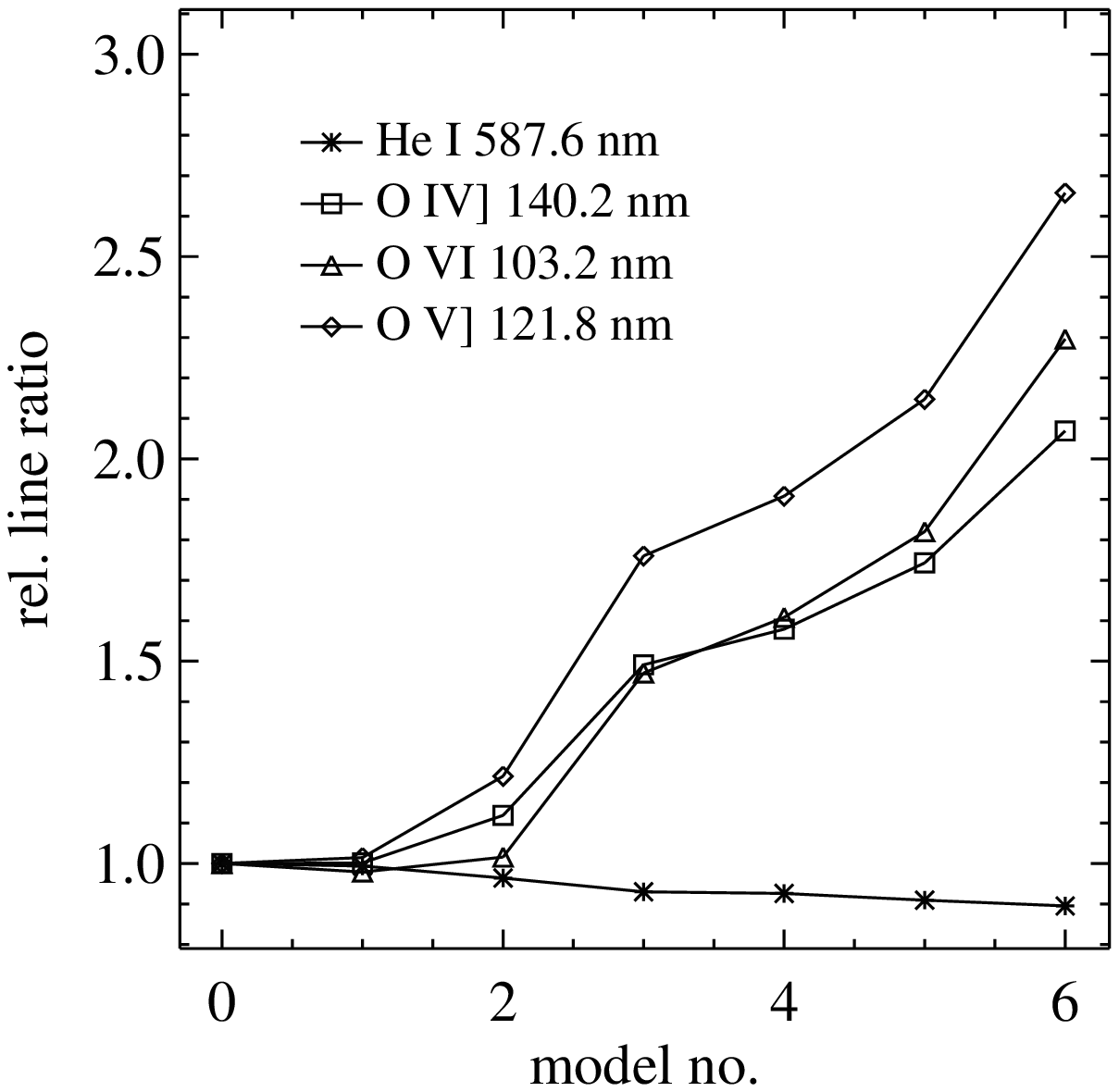}\hfill
\epsfxsize=0.3\textwidth\epsfbox[120 318 450 653]{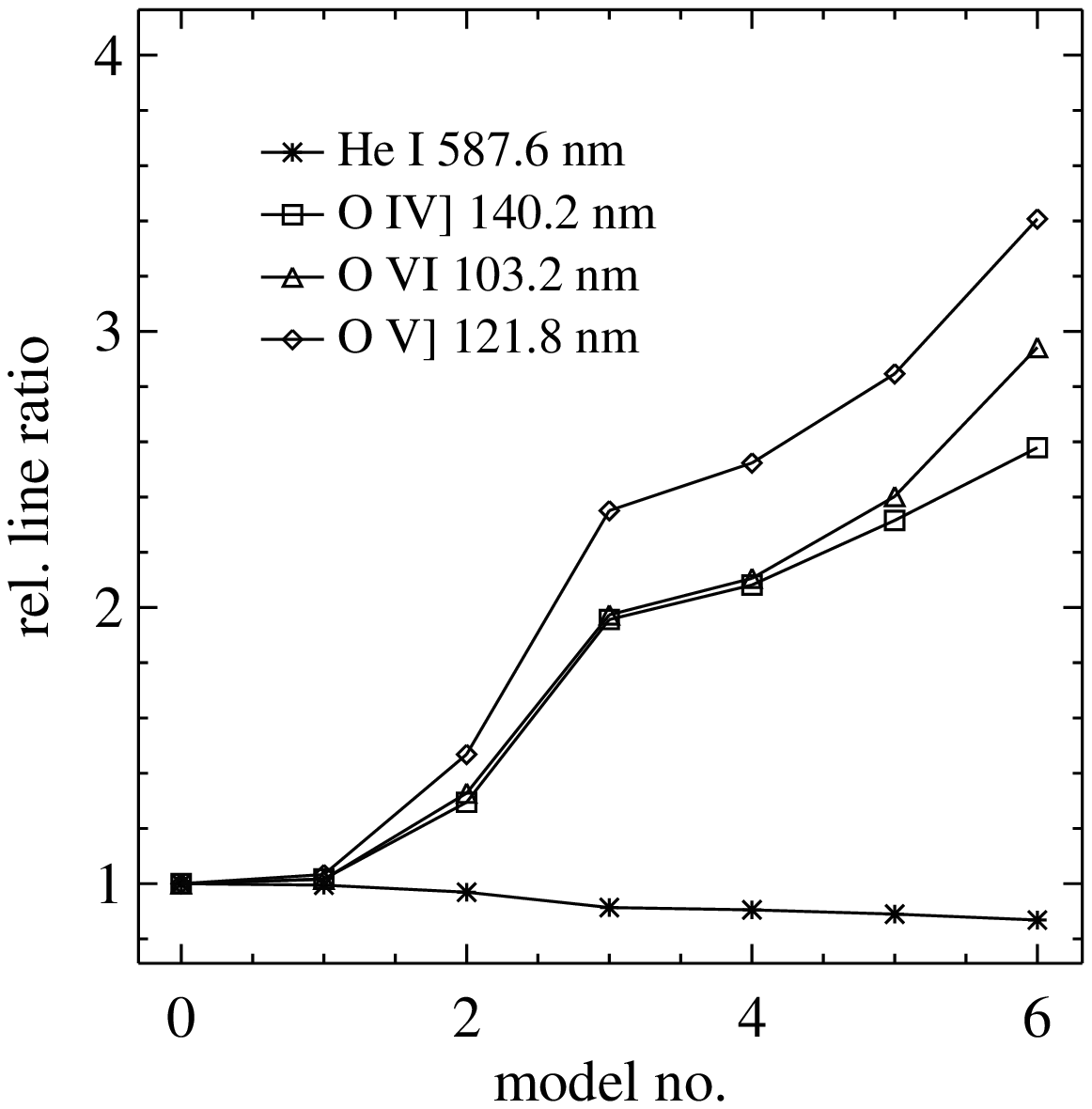}\hfill
}
\caption{The variation of the line ratio to H$\beta$ for selected emission
lines as a function of the grain size distribution. The left panel shows the
results for the H\,{\sc ii}-region models, the middle panel for the PN models
with silicate, and the right panel for the PN models with graphite. All line
flux ratios are normalized to the base model without dust. The models numbers
are defined in Table~\ref{grain:prop}.\label{fig:ratio}}
\end{figure*}

\begin{table*}
\caption{Results for the H\,{\sc ii} region models. The line flux ratio
relative to H$\beta$ = 1 is given for the most important emission lines in the
model, as well as certain relevant physical quantities such as the total
H$\beta$ flux, the electron temperature, helium ionisation fractions, the
outer radius, the continuum optical depth at 121.6~nm, and the fraction of the
total gas heating and cooling contributed by the grains. Quantities between
pointed brackets are volume averages over the entire ionised region.}
\label{hii:tab}
\begin{tabular}{lrrrrrrr}
\hline
 model label     & \ctr{none} & 1.0~$\mu$m & 0.1~$\mu$m & MRN55 & KMH53 & WD55 A0 & WD55 A3 \\
 model no.       & \ctr{0} & \ctr{1} & \ctr{2} & \ctr{3} & \ctr{4} & \ctr{5} & \ctr{6} \\
\hline
 He\,{\sc i} 587.6~nm           & 0.1181 & 0.1167 & 0.1203 & 0.1229 & 0.1301 & 0.1316 & 0.1353 \\ \relax
 C\,{\sc ii}] 232.6~nm          & 0.1995 & 0.2020 & 0.2123 & 0.2137 & 0.2141 & 0.2137 & 0.2041 \\ \relax
 C\,{\sc iii}] 190.9~nm         & 0.0533 & 0.0538 & 0.0862 & 0.1009 & 0.1673 & 0.2162 & 0.4784 \\ \relax
 [N\,{\sc ii}] 658.4~nm         & 0.5331 & 0.5423 & 0.5335 & 0.5188 & 0.4703 & 0.4586 & 0.4247 \\ \relax
 [N\,{\sc ii}] 121.7~$\mu$m     & 0.0281 & 0.0286 & 0.0269 & 0.0257 & 0.0219 & 0.0210 & 0.0184 \\ \relax
 [N\,{\sc iii}] 57.32~$\mu$m    & 0.2981 & 0.2950 & 0.3183 & 0.3292 & 0.3641 & 0.3776 & 0.4162 \\ \relax
 [O\,{\sc ii}] 372.7~nm         & 2.0913 & 2.1158 & 2.2401 & 2.2578 & 2.3174 & 2.3633 & 2.4778 \\ \relax
 [O\,{\sc iii}] 500.7~nm        & 1.5023 & 1.5172 & 2.0366 & 2.2263 & 2.9554 & 3.4175 & 5.0999 \\ \relax
 [O\,{\sc iii}] 51.80~$\mu$m    & 1.2139 & 1.1991 & 1.2755 & 1.3179 & 1.4505 & 1.4979 & 1.6344 \\ \relax
 [O\,{\sc iii}] 88.33~$\mu$m    & 1.1114 & 1.1074 & 1.2076 & 1.2525 & 1.3953 & 1.4523 & 1.6240 \\ \relax
 [Ne\,{\sc ii}] 12.81~$\mu$m    & 0.1765 & 0.1774 & 0.1739 & 0.1699 & 0.1574 & 0.1540 & 0.1424 \\ \relax
 [Ne\,{\sc iii}] 386.9~nm       & 0.0583 & 0.0590 & 0.0842 & 0.0942 & 0.1344 & 0.1610 & 0.2659 \\ \relax
 [Ne\,{\sc iii}] 15.55~$\mu$m   & 0.2924 & 0.2886 & 0.3120 & 0.3255 & 0.3697 & 0.3870 & 0.4401 \\ \relax
 [S\,{\sc ii}] 672.0~nm         & 0.1555 & 0.1579 & 0.1715 & 0.1724 & 0.1694 & 0.1704 & 0.1736 \\ \relax
 [S\,{\sc iii}] 953.1~nm        & 0.7860 & 0.7909 & 0.8462 & 0.8597 & 0.9135 & 0.9466 & 1.0397 \\ \relax
 [S\,{\sc iii}] 18.71~$\mu$m    & 0.6214 & 0.6215 & 0.6218 & 0.6205 & 0.6194 & 0.6207 & 0.6227 \\ \relax
 [S\,{\sc iii}] 33.47~$\mu$m    & 1.2270 & 1.2214 & 1.2203 & 1.2171 & 1.2113 & 1.2120 & 1.2115 \\ \relax
 [S\,{\sc iv}] 10.51~$\mu$m     & 0.5312 & 0.5315 & 0.5876 & 0.6130 & 0.6960 & 0.7340 & 0.8532 \\ \relax
 log[F(H$\beta$)/erg\,s$^{-1}$] & 37.302 & 37.291 & 37.169 & 37.139 & 37.086 & 37.065 & 36.979 \\ \relax
 $T_{\rm e}/$K inner edge       & 7517   & 7597   & 8290   & 8444   & 8805   & 9274   & 11090  \\ \relax
 $<T_{\rm e}/{\rm K}>$          & 7998   & 8017   & 8318   & 8395   & 8670   & 8851   & 9462   \\ \relax
 $<$He$^0$/He$>$                & 0.253  & 0.262  & 0.215  & 0.191  & 0.117  & 0.095  & 0.051  \\ \relax
 $<$He$^+$/He$>$                & 0.746  & 0.738  & 0.785  & 0.809  & 0.883  & 0.904  & 0.948  \\ \relax
 $R_{\rm out}/10^{19}$~cm       & 1.442  & 1.431  & 1.319  & 1.292  & 1.252  & 1.239  & 1.186  \\ \relax
 $\tau_{121.6}$                 & 0.107  & 0.154  & 0.729  & 0.903  & 0.668  & 0.655  & 0.626  \\ \relax
 $<\!\!a^2\!\!>/<\!\!a^3\!\!>$ ($\mu$m$^{-1}$) & --- & 1.00 & 10.00 & 11.55 & 15.33 & 27.54 & 112.95 \\ \relax
 GrGH/TotH                      & 0.000  & 0.005  & 0.059  & 0.072  & 0.130  & 0.185  & 0.348  \\ \relax
 GrGC/TotC                      & 0.000  & 0.0019 & 0.019  & 0.024  & 0.038  & 0.056  & 0.136  \\
\hline
\end{tabular}
\end{table*}

\begin{table*}
\caption{Same as Table~\ref{hii:tab}, but for the planetary nebula region models with silicate.}
\label{pn:sil:tab}
\begin{tabular}{lrrrrrrr}
\hline
 model label     & \ctr{none} & 1.0~$\mu$m & 0.1~$\mu$m & WD31 A0 & WD31 A6 & MRN31 & KMH31 \\
 model no.       & \ctr{0} & \ctr{1} & \ctr{2} & \ctr{3} & \ctr{4} & \ctr{5} & \ctr{6} \\
\hline
 He\,{\sc i} 587.6~nm           & 0.0972 & 0.0966 & 0.0937 & 0.0904 & 0.0900 & 0.0884 & 0.0870 \\ \relax
 He\,{\sc ii} 468.6~nm          & 0.3431 & 0.3470 & 0.3728 & 0.3986 & 0.4018 & 0.4146 & 0.4255 \\ \relax
 C\,{\sc ii}] 232.6~nm          & 0.2826 & 0.2827 & 0.2907 & 0.3084 & 0.3115 & 0.3202 & 0.3293 \\ \relax
 C\,{\sc iii}] 190.9~nm         & 1.7704 & 1.7621 & 1.8008 & 2.0272 & 2.0755 & 2.1789 & 2.3238 \\ \relax
 C\,{\sc iv} 154.9~nm           & 2.4635 & 2.3316 & 1.6057 & 2.1595 & 2.2977 & 2.3985 & 2.8400 \\ \relax
 [N\,{\sc ii}] 658.4~nm         & 0.8455 & 0.8453 & 0.8732 & 0.9147 & 0.9213 & 0.9409 & 0.9619 \\ \relax
 N\,{\sc iii}] 174.9~nm         & 0.0104 & 0.0104 & 0.0107 & 0.0123 & 0.0127 & 0.0135 & 0.0146 \\ \relax
 [N\,{\sc iii}] 57.32~$\mu$m    & 0.1270 & 0.1268 & 0.1254 & 0.1256 & 0.1258 & 0.1257 & 0.1260 \\ \relax
 N\,{\sc iv}] 148.6~nm          & 0.2264 & 0.2264 & 0.2483 & 0.3189 & 0.3349 & 0.3658 & 0.4246 \\ \relax
 N\,{\sc v} 124.0~nm            & 0.1635 & 0.1566 & 0.1306 & 0.1792 & 0.1939 & 0.2118 & 0.2575 \\ \relax
 [O\,{\sc i}] 630.0~nm          & 0.1133 & 0.1127 & 0.1183 & 0.1252 & 0.1260 & 0.1295 & 0.1330 \\ \relax
 [O\,{\sc ii}] 372.7~nm         & 2.0807 & 2.0818 & 2.1505 & 2.2646 & 2.2845 & 2.3396 & 2.3940 \\ \relax
 [O\,{\sc iii}] 436.3~nm        & 0.1599 & 0.1592 & 0.1608 & 0.1773 & 0.1808 & 0.1877 & 0.1975 \\ \relax
 [O\,{\sc iii}] 500.7~nm        & 16.296 & 16.232 & 16.053 & 16.611 & 16.745 & 16.926 & 17.224 \\ \relax
 [O\,{\sc iii}] 51.80~$\mu$m    & 1.2870 & 1.2823 & 1.2528 & 1.2359 & 1.2348 & 1.2244 & 1.2176 \\ \relax
 [O\,{\sc iii}] 88.33~$\mu$m    & 0.2637 & 0.2628 & 0.2568 & 0.2542 & 0.2541 & 0.2523 & 0.2513 \\ \relax
 O\,{\sc iv}] 140.2~nm          & 0.2093 & 0.2096 & 0.2342 & 0.3121 & 0.3305 & 0.3648 & 0.4331 \\ \relax
 [O\,{\sc iv}] 25.88~$\mu$m     & 3.6176 & 3.6370 & 3.8972 & 4.2203 & 4.2682 & 4.4150 & 4.5757 \\ \relax
 O\,{\sc v}] 121.8~nm           & 0.1836 & 0.1863 & 0.2232 & 0.3232 & 0.3503 & 0.3943 & 0.4879 \\ \relax
 O\,{\sc vi} 103.2~nm           & 0.0189 & 0.0185 & 0.0192 & 0.0278 & 0.0304 & 0.0344 & 0.0434 \\ \relax
 [Ne\,{\sc ii}] 12.81~$\mu$m    & 0.0250 & 0.0249 & 0.0255 & 0.0260 & 0.0260 & 0.0264 & 0.0266 \\ \relax
 [Ne\,{\sc iii}] 386.9~nm       & 2.0600 & 2.0536 & 2.0574 & 2.1590 & 2.1807 & 2.2197 & 2.2741 \\ \relax
 [Ne\,{\sc iii}] 15.55~$\mu$m   & 1.8676 & 1.8475 & 1.8191 & 1.8027 & 1.8019 & 1.7923 & 1.7868 \\ \relax
 [Ne\,{\sc iv}] 242.3~nm        & 0.7983 & 0.8008 & 0.8748 & 1.0612 & 1.0989 & 1.1801 & 1.3167 \\ \relax
 [Ne\,{\sc v}] 342.6~nm         & 0.5785 & 0.5834 & 0.6429 & 0.7661 & 0.7919 & 0.8436 & 0.9286 \\ \relax
 [Ne\,{\sc v}] 14.32~$\mu$m     & 1.5680 & 1.5787 & 1.6792 & 1.7906 & 1.8040 & 1.8569 & 1.9041 \\ \relax
 [Ne\,{\sc v}] 24.31~$\mu$m     & 1.0480 & 1.0513 & 1.1239 & 1.2128 & 1.2251 & 1.2664 & 1.3088 \\ \relax
 [Ne\,{\sc vi}] 7.65~$\mu$m     & 0.1100 & 0.1112 & 0.1212 & 0.1341 & 0.1361 & 0.1423 & 0.1491 \\ \relax
 Mg\,{\sc ii} 279.8~nm          & 2.2299 & 2.1744 & 2.2295 & 2.3249 & 2.3440 & 2.3828 & 2.4212 \\ \relax
 [Mg\,{\sc iv}] 4.485~$\mu$m    & 0.1224 & 0.1236 & 0.1316 & 0.1407 & 0.1420 & 0.1459 & 0.1502 \\ \relax
 [Mg\,{\sc v}] 292.8~nm         & 0.0937 & 0.0941 & 0.1020 & 0.1202 & 0.1240 & 0.1313 & 0.1441 \\ \relax
 [Mg\,{\sc v}] 5.608~$\mu$m     & 0.1848 & 0.1857 & 0.1942 & 0.2052 & 0.2067 & 0.2112 & 0.2163 \\ \relax
 Si\,{\sc ii}] 233.5~nm         & 0.1647 & 0.1640 & 0.1667 & 0.1776 & 0.1796 & 0.1848 & 0.1905 \\ \relax
 [Si\,{\sc ii}] 34.81~$\mu$m    & 0.1566 & 0.1561 & 0.1611 & 0.1674 & 0.1681 & 0.1713 & 0.1743 \\ \relax
 Si\,{\sc iii}] 189.2~nm        & 0.4584 & 0.4624 & 0.5051 & 0.5949 & 0.6132 & 0.6583 & 0.7194 \\ \relax
 Si\,{\sc iv} 139.7~nm          & 0.2058 & 0.1960 & 0.1457 & 0.1825 & 0.1936 & 0.2047 & 0.2400 \\ \relax
 [S\,{\sc ii}] 672.0~nm         & 0.3698 & 0.3698 & 0.3823 & 0.4011 & 0.4038 & 0.4129 & 0.4222 \\ \relax
 [S\,{\sc iii}] 953.1~nm        & 1.1365 & 1.1356 & 1.1529 & 1.1918 & 1.1991 & 1.2162 & 1.2346 \\ \relax
 [S\,{\sc iii}] 18.71~$\mu$m    & 0.5437 & 0.5435 & 0.5520 & 0.5628 & 0.5644 & 0.5694 & 0.5743 \\ \relax
 [S\,{\sc iii}] 33.47~$\mu$m    & 0.2180 & 0.2178 & 0.2214 & 0.2269 & 0.2278 & 0.2303 & 0.2329 \\ \relax
 [S\,{\sc iv}] 10.51~$\mu$m     & 2.0387 & 2.0307 & 1.9690 & 1.9321 & 1.9296 & 1.9119 & 1.9012 \\ \relax
 log[F(H$\beta$)/erg\,s$^{-1}$] & 35.432 & 35.424 & 35.361 & 35.317 & 35.313 & 35.288 & 35.271 \\ \relax
 $T_{\rm e}/$K inner edge       & 17990  & 18040  & 18750  & 19910  & 20210  & 20500  & 21210  \\ \relax
 $<T_{\rm e}/{\rm K}>$          & 12050  & 12050  & 12190  & 12560  & 12618  & 12764  & 12972  \\ \relax
 $<$He$^+$/He$>$                & 0.670  & 0.665  & 0.643  & 0.618  & 0.615  & 0.604  & 0.592  \\ \relax        
 $<$He$^{2+}$/He$>$             & 0.297  & 0.300  & 0.320  & 0.343  & 0.346  & 0.356  & 0.368  \\ \relax
 $R_{\rm out}/10^{17}$~cm       & 4.059  & 4.033  & 3.863  & 3.768  & 3.760  & 3.705  & 3.672  \\ \relax
 $\tau_{121.6}$                 & 0.267  & 0.274  & 0.404  & 0.504  & 0.489  & 0.566  & 0.558  \\ \relax
 $<\!\!a^2\!\!>/<\!\!a^3\!\!>$ ($\mu$m$^{-1}$) & --- & 1.00 & 10.00 & 25.08 & 38.44 & 28.28 & 41.54 \\ \relax
 GrGH/TotH                      & 0.000  & 0.0013 & 0.017  & 0.050  & 0.060  & 0.066  & 0.091  \\ \relax
 GrGC/TotC                      & 0.000  & 0.0005 & 0.005  & 0.014  & 0.019  & 0.017  & 0.024  \\
\hline
\end{tabular}
\end{table*}

\begin{table*}
\caption{Same as Table~\ref{hii:tab}, but for the planetary nebula region models with graphite.}
\label{pn:gra:tab}
\begin{tabular}{lrrrrrrr}
\hline
 model label     & \ctr{none} & 1.0~$\mu$m & 0.1~$\mu$m & MRN31 & WD31 A0 & KMH31 & WD31 A6 \\
 model no.       & \ctr{0} & \ctr{1} & \ctr{2} & \ctr{3} & \ctr{4} & \ctr{5} & \ctr{6} \\
\hline
 He\,{\sc i} 587.6~nm           & 0.0972 & 0.0967 & 0.0942 & 0.0888 & 0.0880 & 0.0865 & 0.0844 \\ \relax
 He\,{\sc ii} 468.6~nm          & 0.3431 & 0.3459 & 0.3665 & 0.4089 & 0.4147 & 0.4272 & 0.4463 \\ \relax
 C\,{\sc ii}] 232.6~nm          & 0.2826 & 0.2827 & 0.2915 & 0.3240 & 0.3294 & 0.3407 & 0.3427 \\ \relax
 C\,{\sc iii}] 190.9~nm         & 1.7704 & 1.7663 & 1.8675 & 2.3375 & 2.4196 & 2.5850 & 2.6638 \\ \relax
 C\,{\sc iv} 154.9~nm           & 2.4635 & 2.3414 & 1.8614 & 2.5680 & 2.8158 & 2.8801 & 3.3301 \\ \relax
 [N\,{\sc ii}] 658.4~nm         & 0.8455 & 0.8449 & 0.8709 & 0.9397 & 0.9515 & 0.9721 & 0.9906 \\ \relax
 N\,{\sc iii}] 174.9~nm         & 0.0104 & 0.0104 & 0.0112 & 0.0147 & 0.0153 & 0.0165 & 0.0172 \\ \relax
 [N\,{\sc iii}] 57.32~$\mu$m    & 0.1270 & 0.1269 & 0.1265 & 0.1273 & 0.1274 & 0.1276 & 0.1265 \\ \relax
 N\,{\sc iv}] 148.6~nm          & 0.2264 & 0.2292 & 0.2798 & 0.4079 & 0.4315 & 0.4755 & 0.5216 \\ \relax
 N\,{\sc v} 124.0~nm            & 0.1635 & 0.1600 & 0.1585 & 0.2469 & 0.2685 & 0.2922 & 0.3582 \\ \relax
 [O\,{\sc i}] 630.0~nm          & 0.1133 & 0.1127 & 0.1175 & 0.1286 & 0.1303 & 0.1337 & 0.1364 \\ \relax
 [O\,{\sc ii}] 372.7~nm         & 2.0807 & 2.0807 & 2.1479 & 2.3458 & 2.3791 & 2.4430 & 2.4695 \\ \relax
 [O\,{\sc iii}] 436.3~nm        & 0.1599 & 0.1595 & 0.1662 & 0.2014 & 0.2072 & 0.2183 & 0.2226 \\ \relax
 [O\,{\sc iii}] 500.7~nm        & 16.296 & 16.252 & 16.354 & 17.608 & 17.806 & 18.134 & 18.087 \\ \relax
 [O\,{\sc iii}] 51.80~$\mu$m    & 1.2870 & 1.2832 & 1.2639 & 1.2415 & 1.2387 & 1.2307 & 1.2106 \\ \relax
 [O\,{\sc iii}] 88.33~$\mu$m    & 0.2637 & 0.2631 & 0.2593 & 0.2564 & 0.2561 & 0.2550 & 0.2510 \\ \relax
 O\,{\sc iv}] 140.2~nm          & 0.2093 & 0.2129 & 0.2710 & 0.4095 & 0.4357 & 0.4846 & 0.5397 \\ \relax
 [O\,{\sc iv}] 25.88~$\mu$m     & 3.6176 & 3.6385 & 3.8995 & 4.4292 & 4.5055 & 4.6592 & 4.8556 \\ \relax
 O\,{\sc v}] 121.8~nm           & 0.1836 & 0.1897 & 0.2696 & 0.4316 & 0.4633 & 0.5226 & 0.6255 \\ \relax
 O\,{\sc vi} 103.2~nm           & 0.0189 & 0.0192 & 0.0251 & 0.0373 & 0.0398 & 0.0454 & 0.0556 \\ \relax
 [Ne\,{\sc ii}] 12.81~$\mu$m    & 0.0250 & 0.0250 & 0.0255 & 0.0261 & 0.0261 & 0.0264 & 0.0265 \\ \relax
 [Ne\,{\sc iii}] 386.9~nm       & 2.0600 & 2.0556 & 2.0899 & 2.2993 & 2.3338 & 2.3926 & 2.4068 \\ \relax
 [Ne\,{\sc iii}] 15.55~$\mu$m   & 1.8676 & 1.8571 & 1.8392 & 1.8149 & 1.8123 & 1.8046 & 1.7876 \\ \relax
 Ne\,{\sc iv}] 242.3~nm         & 0.7983 & 0.8069 & 0.9438 & 1.2677 & 1.3235 & 1.4304 & 1.5390 \\ \relax
 [Ne\,{\sc v}] 342.6~nm         & 0.5785 & 0.5869 & 0.6896 & 0.8825 & 0.9130 & 0.9738 & 1.0497 \\ \relax
 [Ne\,{\sc v}] 14.32~$\mu$m     & 1.5680 & 1.5764 & 1.6765 & 1.8595 & 1.8800 & 1.9344 & 1.9913 \\ \relax
 [Ne\,{\sc v}] 24.31~$\mu$m     & 1.0480 & 1.0524 & 1.1328 & 1.2769 & 1.2943 & 1.3377 & 1.3833 \\ \relax
 [Ne\,{\sc vi}] 7.65~$\mu$m     & 0.1100 & 0.1111 & 0.1230 & 0.1424 & 0.1447 & 0.1510 & 0.1581 \\ \relax
 Mg\,{\sc ii} 279.8~nm          & 2.2299 & 2.2030 & 1.9523 & 1.9529 & 2.0220 & 1.9956 & 2.0640 \\ \relax
 [Mg\,{\sc iv}] 4.485~$\mu$m    & 0.1224 & 0.1234 & 0.1311 & 0.1459 & 0.1479 & 0.1521 & 0.1572 \\ \relax
 [Mg\,{\sc v}] 292.8~nm         & 0.0937 & 0.0948 & 0.1102 & 0.1388 & 0.1432 & 0.1519 & 0.1616 \\ \relax
 [Mg\,{\sc v}] 5.608~$\mu$m     & 0.1848 & 0.1856 & 0.1965 & 0.2135 & 0.2153 & 0.2200 & 0.2238 \\ \relax
 Si\,{\sc ii}] 233.5~nm         & 0.1647 & 0.1640 & 0.1674 & 0.1884 & 0.1922 & 0.1987 & 0.2034 \\ \relax
 [Si\,{\sc ii}] 34.81~$\mu$m    & 0.1566 & 0.1561 & 0.1604 & 0.1707 & 0.1723 & 0.1749 & 0.1809 \\ \relax
 Si\,{\sc iii}] 189.2~nm        & 0.4584 & 0.4627 & 0.5224 & 0.7002 & 0.7289 & 0.7982 & 0.8331 \\ \relax
 Si\,{\sc iv} 139.7~nm          & 0.2058 & 0.1980 & 0.1694 & 0.2631 & 0.2893 & 0.3080 & 0.3584 \\ \relax
 [S\,{\sc ii}] 672.0~nm         & 0.3698 & 0.3697 & 0.3808 & 0.4119 & 0.4170 & 0.4262 & 0.4348 \\ \relax
 [S\,{\sc iii}] 953.1~nm        & 1.1365 & 1.1356 & 1.1562 & 1.2272 & 1.2389 & 1.2599 & 1.2712 \\ \relax
 [S\,{\sc iii}] 18.71~$\mu$m    & 0.5437 & 0.5435 & 0.5522 & 0.5700 & 0.5725 & 0.5773 & 0.5814 \\ \relax
 [S\,{\sc iii}] 33.47~$\mu$m    & 0.2180 & 0.2178 & 0.2216 & 0.2310 & 0.2325 & 0.2351 & 0.2373 \\ \relax
 [S\,{\sc iv}] 10.51~$\mu$m     & 2.0387 & 2.0334 & 1.9881 & 1.9380 & 1.9325 & 1.9210 & 1.8964 \\ \relax
 log[F(H$\beta$)/erg\,s$^{-1}$] & 35.432 & 35.425 & 35.378 & 35.318 & 35.313 & 35.294 & 35.275 \\ \relax
 $T_{\rm e}/$K inner edge       & 17990  & 18110  & 19320  & 20500  & 20710  & 21000  & 21920  \\ \relax
 $<T_{\rm e}/{\rm K}>$          & 12050  & 12078  & 12303  & 12912  & 13002  & 13183  & 13335  \\ \relax
 $<$He$^+$/He$>$                & 0.670  & 0.665  & 0.644  & 0.605  & 0.598  & 0.587  & 0.569  \\ \relax
 $<$He$^{2+}$/He$>$             & 0.297  & 0.299  & 0.319  & 0.355  & 0.360  & 0.371  & 0.384  \\ \relax
 $R_{\rm out}/10^{17}$~cm       & 4.059  & 4.039  & 3.919  & 3.799  & 3.793  & 3.753  & 3.718  \\ \relax
 $\tau_{121.6}$                 & 0.267  & 0.273  & 0.388  & 0.502  & 0.469  & 0.512  & 0.437  \\ \relax
 $<\!\!a^2\!\!>/<\!\!a^3\!\!>$ ($\mu$m$^{-1}$) & --- & 1.00 & 10.00 & 28.28 & 61.10 & 43.31 & 326.53 \\ \relax
 GrGH/TotH                      & 0.000  & 0.003  & 0.038  & 0.090  & 0.103  & 0.114  & 0.163  \\ \relax
 GrGC/TotC                      & 0.000  & 0.0005 & 0.006  & 0.017  & 0.028  & 0.025  & 0.086  \\
\hline			        
\end{tabular}
\end{table*}

In Tables~\ref{hii:tab}, \ref{pn:sil:tab}, and \ref{pn:gra:tab} we present
the results of these calculations. These tables contain a comparison of the
flux (presented as a ratio to H$\beta$=1 for that particular model) of the
most important infrared fine-structure lines, as well as optical and UV
emission lines. A number of physical parameters are also compared, such as the
total H$\beta$ flux, the electron temperature at the inner edge of the nebula
as well as the volume average over the entire nebula, the fractional ionization of helium
(given as the ratio of the ion abundance over the total helium abundance), the
outer radius, and the fraction that the photo-electric heating and collisional
cooling contribute to the total heating and cooling of the gas.

The WD55 A3 and WD31 A6 size distributions contain enhanced amounts of very
small carbonaceous particles (PAHs). Such particles are not expected to exist
in ionized regions, and hence the models with graphite based on these size
distributions are less realistic than those based on other size distributions.
In the following discussion, results pertaining to these models will only be
included in parenthesis.

The numbers in Tables~\ref{hii:tab}, \ref{pn:sil:tab}, and \ref{pn:gra:tab}
show that adding a dust component has a dramatic effect on the physical
conditions and the normalized spectrum. In the H\,{\sc ii} region models the
strongest enhancements in line strength (compared to the base model without
dust) are for C\,{\sc iii}] 190.9~nm with a factor 4.06 enhancement for the
WD55 A0 size distribution (WD55 A3: 8.98), followed by [Ne\,{\sc iii}]
386.9~nm with a factor 2.76 (4.56), and [O\,{\sc iii}] 500.7~nm with a factor
2.27 (3.39). The strongest decrease is for the [N\,{\sc ii}] 121.7~$\mu$m line
with a factor 0.747 (WD55 A3: 0.655), followed by [N\,{\sc ii}] 658.4~nm with
a factor 0.860 (0.797), and [Ne\,{\sc ii}] 12.81~$\mu$m with a factor 0.873
(0.807). For the PN models the lines with the strongest enhancement are
O\,{\sc v}] 121.8~nm with a factor 2.66 for silicate and 2.85 for graphite
(WD31 A6: 3.41), followed by O\,{\sc vi} 103.2~nm with factors 2.30 for
silicate and 2.40 (2.94) for graphite, and O\,{\sc iv}] 140.2~nm with factors
2.07 for silicate and 2.32 (2.58) for graphite. The strongest decrease in the
silicate models is for the He\,{\sc i} 587.6~nm line with a factor 0.895,
followed by [S\,{\sc iv}] 10.51~$\mu$m with a factor 0.933, and [O\,{\sc iii}]
51.80~$\mu$m with a factor 0.946. The strongest decrease in the graphite
models is for the He\,{\sc i} 587.6~nm line with a factor 0.890 (WD31 A6:
0.868), followed by Mg\,{\sc ii} 279.8~nm with a factor 0.895 (0.926), and
[S\,{\sc iv}] 10.51~$\mu$m with a factor 0.942 (0.930). The variation of the
relative strength of these lines is also shown in Fig.~\ref{fig:ratio}.

What is apparent from this comparison is that the strongest enhancement is
always found in optical/UV lines of the highest ionization stages
in the spectrum,
while the strongest decrease is typically found in optical/UV
lines of low ionization lines or infrared fine-structure lines of
low/intermediate ionization stages. This indicates that the effect is
strongest in the inner regions of the nebula. This is confirmed by comparing
the electron temperature at the inner edge of the nebula and the average over
the entire volume. It is clear that in all cases the former changes far more
strongly than the latter. What is also clear from the comparison is that the
enhancement in emission line strength correlates well with the total amount of
photo-electric heating contributed to the plasma by the grains. This is no
surprise since most of the emission lines (with the exception of the helium
lines) are predominantly collisionally excited. Small grains contribute far
more to photo-electric heating than large grains, due to their much higher
opacities (see Figs.~\ref{plot:sd:sil} and \ref{plot:sd:gra}) and
photoelectric yields (see Figure 5 in WD). Thus, one can roughly state that
the more small particles a size distribution contains, the stronger the
photo-electric heating of the plasma by the grains will be, which in turn
leads to a stronger effect on the emitted spectrum. The fact that this effect
is strongest at the inner edge can be understood by looking at
Figs.~\ref{plot:sd:sil} and \ref{plot:sd:gra}. It is apparent that for the
smallest grains the absorption opacity peaks at a wavelength that corresponds
fairly well with the Lyman edge of hydrogen. This implies that at the inner
edge the grains will absorb lots of ionizing photons. However, when one moves
away from the illuminated face of the nebula, more and more ionizing
photons will be absorbed away, while non-ionizing photons will be less
affected. Since grains can absorb both, the relative fraction of non-ionizing
photons that the grains absorb will increase. Hence the average energy per
absorbed photon will decrease. For the gas the story is very different. In a
first approximation the electron temperature will be constant inside the
ionized region, and therefore the recombination rates will be constant as
well. Since the gas is assumed to be in equilibrium, the photo-ionization
rates must be constant as well and the total amount of heating is constant too
(or even increases near the ionization front because the lowest energy
ionizing photons will be depleted). This implies that the relative importance
of the photo-electric effect will decrease when one moves away from the
illuminated face of the nebula.

We also noted that the enhancement effect is stronger for optical/UV lines,
compared to infrared fine-structure lines. This is easy to understand. The
excitation potential of the optical/UV lines is much higher and they are only
excited by the high energy tail of the electrons. This
makes these lines exponentially sensitive to electron temperature, and
therefore also very sensitive to the enhanced photo-electric heating. The
infrared fine-structure lines on the other hand have low excitation potentials
and are nearly insensitive to electron temperature and photo-electric heating.
Most of the effect on these lines will be due to changes in the overall
ionization structure. A mild temperature dependence of the collisional cross
section caused by resonances may also contribute (e.g., [Ne V] 14.32~$\mu$m,
see van Hoof et al. 2000a and references therein).

From a comparison of the models it is also apparent that the helium
recombination lines are affected by the photo-electric effect. This has
already been explained earlier. The increased photo-electric effect leads to
an increase in electron temperature, which in turn leads to a decrease in the
recombination rates and an increase in the overall degree of ionization. This
is confirmed by comparing the volume averages of the helium ion fractions. It
is clear that in the H\,{\sc ii}-region models He$^+$ is gaining at the
expense of He$^0$, while in the PN models He$^{2+}$ is gaining at the expense
of He$^+$ when the photo-electric effect increases.

\begin{table}
\begin{center}
\caption{Line-centre optical depths for the resonance lines in the PN model
with 0.1~$\mu$m silicate grains. All the resonance lines are doublets, which
are shown separately here.}
\label{opt:dep}
\begin{tabular}{lr}
\hline
line & opt. depth \\
\hline
 C\,{\sc iv} 154.8~nm  & 6760 \\
 C\,{\sc iv} 155.1~nm  & 3390 \\
 Si\,{\sc iv} 139.4~nm & 2510 \\
 Si\,{\sc iv} 140.3~nm & 1260 \\
 N\,{\sc v} 123.9~nm   & 1030 \\
 N\,{\sc v} 124.3~nm   &  512 \\
 Mg\,{\sc ii} 279.6~nm & 2790 \\
 Mg\,{\sc ii} 280.4~nm & 1400 \\
 O\,{\sc vi} 103.2~nm  &  517 \\
 O\,{\sc vi} 103.8~nm  &  258 \\
\hline
\end{tabular}
\end{center}
\end{table}

In general the relative line strength increases or decreases monotonically
with an increase in the photo-electric effect. However, there are a couple of
notable exceptions to this rule in the PN models. In the following discussion
we will concentrate on the silicate models. The details are somewhat different
for the graphite models, but the underlying physics is the same. The
exceptional behaviour occurs for the following lines (in order of decreasing
magnitude of the effect): C\,{\sc iv} 154.9~nm, Si\,{\sc iv} 139.7~nm, and
N\,{\sc v} 124.0~nm. For all these lines the relative strength decreases going
from the dust-free model to the 1.0~$\mu$m and the 0.1~$\mu$m models, while
showing an overall increasing trend for the subsequent models. What is
immediately apparent is that this effect only occurs in allowed (resonance)
transitions. However, the effect is much less in the other two resonance lines
in the list, viz.\ Mg\,{\sc ii} 279.8~nm and O\,{\sc vi} 103.2~nm (these lines
only show a mild decrease going from the dust-free to the 1.0~$\mu$m model).
All five lines are the equivalent of Ly$\alpha$. They have large
line-centre optical depths, and as a consequence they are scattered
many times in the nebula before escaping.
This random walk
greatly increases the chance of the photon being absorbed by background
opacities, most notably from dust grains. The higher the line-centre optical
depth is, the longer the random walk will be and the higher the probability of
destruction is. The fact that the strongest decrease occurs going from the
1.0~$\mu$m to the 0.1~$\mu$m model can be understood by looking at
Fig.~\ref{plot:sd:sil}. The dust opacity rises dramatically in the UV going
from 1.0~$\mu$m to 0.1~$\mu$m-sized grains, while the photo-electric effect is
still modest for these grains. Hence the increase in photon destruction wins.
For subsequent models the rise in optical depth is less dramatic (or there is
even a decrease), while the photo-electric effect starts to dominate the
models. In these models the rising photo-electric effect wins. This trend is
confirmed by inspecting $\tau_{121.6}$ in Tables~\ref{hii:tab},
\ref{pn:sil:tab}, and \ref{pn:gra:tab}.\footnote{Note that $\tau_{121.6}$ is
the total continuum optical depth due to all physical processes, corrected
for stimulated emission. This implies that the optical depth should be non-zero
for the dust-free models.} In Table~\ref{opt:dep} we show the
line-centre optical depths of the lines in question. All optical depths follow
the trend of the decreasing magnitude of the effect indicated above, with the
exception of Mg\,{\sc ii} 279.8~nm. For this line the decreasing effect is
much weaker than expected based on the line-centre optical depth. This line
has a much longer wavelength than any of the other lines. At 279.8~nm silicate
grains are more or less transparent and the chance of absorption on grains is
greatly reduced due to this fact. This is not the case for graphite grains,
and one can see that the decreasing effect is stronger in the graphite models
for this line.

From the discussion presented above, one might be tempted to say that the
photo-electric effect will be stronger if more small grains are present, due
to the increased opacity and the higher photo-electric yield. In a rough sense
this is true, but the reader should be warned that no simple predictors can be
constructed from such an observation that allow an accurate estimate of the
effect.
Examples of such predictors could be the average surface to
volume ratio $<\!\!a^2\!\!>/<\!\!a^3\!\!>$ of the grains and the continuum
optical depth $\tau_{121.6}$. It can be seen from Tables~\ref{hii:tab},
\ref{pn:sil:tab}, and \ref{pn:gra:tab} that in general neither of these
predict the correct sequence in the tables. Detailed models will always be
necessary to obtain a reliable prediction of the magnitude of the
photo-electric effect.


All these results clearly illustrate that the size distribution alone has a
dramatic effect on the emitted spectrum, and is therefore an important
parameter in the modelling of spectra from H\,{\sc ii} regions and PNe.
However, very little is known about the size distribution of grains in these
objects. As was already mentioned above, the most detailed studies of grain
size distributions in the literature focus on explaining the extinction curve
caused by grains in the diffuse ISM. It is not clear whether these
distributions are valid for H\,{\sc ii} regions since grains undergo an
appreciable amount of processing when they move in and out of molecular
clouds.
It is even more questionable
whether ISM size distributions would be valid for PN. After all, the grains in
the ISM come from a variety of sources (including PN, but also supernovae and
other sources), and also the grains in the ISM have undergone far more
prolonged processing than the grains in the PN. Therefore the further study of
grain size distributions in photo-ionized regions, as well as AGB/post-AGB
stars is urgently needed.

\section{Conclusions}
\label{conclusions}

In this paper we investigated the effect the grain size distribution has on
the amount of photo-electric heating in photo-ionized regions, and its
consequences for the spectrum emitted by the plasma. To model these effects we
used the comprehensively upgraded grain code in Cloudy 96. One of these
upgrades is the newly developed hybrid grain charge model. This model allows
discrete charge states of very small grains to be modelled accurately while
simultaneously avoiding the overhead of fully resolving the charge
distribution of large grains, thus making the model both accurate and
computationally efficient.

A comprehensive comparison with the fully resolved charge state models of WD
validates the new model. The WD and Cloudy results for the photo-electric
heating rates are generally in excellent agreement, with only a few outliers
for single sized grains in the $n=2$ and $n=3$ cases. The results for the size
distribution cases always agree to better than 25\%, even for $n=2$. This is
well within the accuracy with which we know grain physics to date. The
collisional cooling rates usually are in even better agreement. It is
furthermore clear that the accuracy of the $n$-charge state approximation
increases as the number of charge states increases, as should be expected. The
agreement between the Cloudy and the WD results is very satisfactory for
realistic size distributions, and should be sufficient for all realistic
astrophysical applications, even with $n=2$.

The effect of the grain size distribution on the line emission from
photo-ionized regions is studied by taking standard models for an H\,{\sc ii}
region and a planetary nebula and adding a dust component to the models with
varying grain size distributions (either single sized grains or size
distributions taken from the literature). A comparison of the models shows
that varying the size distribution (while keeping the chemical composition and
the dust-to-gas mass ratio of the grains constant) has a dramatic effect on
the emitted spectrum. The strongest enhancement is always found in optical/UV
lines of the highest ionization stages present in the spectrum (with factors
up to 2.5 -- 4), while the strongest decrease is typically found in optical/UV
lines of low ionization lines or infrared fine-structure lines of
low/intermediate ionization stages (with reductions up to 10 -- 25\%). The
enhancement effect is strongest in the inner regions of the nebula, and it
correlates well with the total amount of photo-electric heating contributed to
the plasma by the grains. Changing the grain size distribution also affects
the ionization balance because the increase in electron temperature leads to
enhanced recombination and also because the grains are directly competing with
the gas for ionizing photons. Finally, changing the grain size distribution
can also affect resonance lines like C\,{\sc iv} 154.9~nm which are very
sensitive to changes in the background opacity.

All these results clearly demonstrate that the grain size distribution is an
important parameter in photo-ionization models. Only few studies of grain size
distributions exist, and they mainly concentrate on the diffuse interstellar
medium (ISM) in order to explain extinction curves. Further study of grain
size distributions will be needed in order to enable more accurate modelling of
photo-ionized regions. This is especially the case for planetary nebulae since
it is not clear whether ISM size distributions are valid for these objects.

\section*{Acknowledgments}

We wish to acknowledge financial support by the National Science Foundation
through grant no.\ AST--0071180, and NASA through its LTSA program, NAG
5--3223. This research was also supported in part by the Natural Sciences and
Engineering Research Council of Canada. P.v.H. is currently supported by the
Engineering and Physical Sciences Research Council of the United Kingdom.
J.C.W. acknowledges support from an NSF International Research Fellowship.
This paper used the photo-ionization code Cloudy which can be obtained from
{\verb#http://www.nublado.org#}, as well as the Atomic Line List available
at {\verb#http://www.pa.uky.edu/~peter/atomic#}.

\end{document}